\documentclass[11pt]{article} 
\usepackage{amsmath} 
\usepackage{amsthm} 
\usepackage{amssymb}	
 \usepackage{indentfirst}
\usepackage{float}
\usepackage{titlesec}
\titlelabel{\thetitle. \quad}

\usepackage{authblk}
\usepackage{graphicx} 
\usepackage[dvips,letterpaper,margin=0.75in,bottom=0.5in]{geometry}
\usepackage[titletoc,toc,title]{appendix}
\usepackage{verbatim}
\usepackage{cite}
\usepackage{caption}
\usepackage[figurename=FIG.]{caption}
\begin{document}
\title{Quark-Photon-Quark Correlation and Transverse Target Single Spin Asymmetry in Inclusive DIS}

\author{Matthias Burkardt and Tareq Alhalholy }
\affil{Department of Physics, New Mexico State University,
Las Cruces, NM 88003-0001, U.S.A.}
\date{\today}
\maketitle

\begin{abstract}
We calculate the $q \gamma q$ correlation function associated to transverse target lepton-nucleon inclusive deep-inelastic scattering by direct evaluation of the corresponding  matrix element. We use the electromagnetic impact parameter potential for a transversely polarized nucleon to define the $q \gamma q$ correlation function for each flavor in terms of the field components of that flavor. The results are compared with the two existing models for the $q \gamma q$ correlator. Using the calculated $q \gamma q$ correlation function, we estimated the transverse target SSA in IDIS process.

\end{abstract}
\titlepage
\section{Introduction}
Single spin asymmetry in inclusive and semi inclusive deep inelastic scattering is of a prime importance in studying and  revealing the spin structure of the nucleon. In particular, asymmetries appear in inclusive DIS are important to study quark correlation  $q\gamma q $  in the nucleon, while in semi-inclusive DIS the asymmetries give information about the quark-gluon correlation  $qgq$  in the nucleon. Due to its importance in transverse target single spin aymetry,   the $qgq$ correlator is extensively studied~\cite{Qiu91, Qiu98,Shmidt2012, Q-g-correlator}, and extraction from data is done either directly from $pp^{\uparrow}$ collision~\cite{Couvaris} or by extracting the Sivers function from SIDIS data where the $qgq$  correlator is proportional to the first moment of Sivers function~\cite{Boer2015}. 
\newline

\noindent On the other hand, the $q \gamma q $  correlator (associated to a transversely polarized target in inclusive DIS) is modeled by relating it with the  Qiu-Sterman function (or the Efremov-Teryaev-Qiu-Sterman (ETQS) twist-3 matrix element), by rescaling the  $q g q $ correlator~\cite{Metz1,BurkardtFFT} utilizing the symmetry between the  corresponding diagrams reoresenting the two correlators. In the two existing models for the $q \gamma q$ correlator, there are disagreements in sign and magnitude in the  correlators for the majority flavors (up quarks in the proton and down quarks in the neutron), while they agree in sign and magnitude for the proton and neutron minority flavors. In Ref.~\cite{Tareq1}, the electromagnetic potential corresponding to a transversely polarized nucleon was calculated, this potential can be used to explicitly calculate the $q \gamma q$ correlator. In this paper, we present a direct calculation of the  $q \gamma q $  correlator by evaluating the corresponding matrix element using the transverse (impact parameter) electromagnetic potential for a transversely polarized nucleon. The results are then  compared with the existing models for the  $q\gamma q $ correlator.
\section{Transverse charge Density and Impact Parameter Potential for Transversely Polarized Nucleon}
For a proton transversely polarized (with respect to the lepton plane), the transverse charge density (or the impact parameter dependent parton distribution function) is given by~\cite{BurkardtGPDs}
\begin{equation}
\rho_T (x,\mathbf{b_\perp})_{proton} =  \int \frac{d^2 q_{\perp}}{(2 \pi)^2} e^{- i \mathbf{q_\perp \cdot b_\perp}} \left(  \sum_f  e_f \ H^f _v (x,-q^2_\perp) \ + \ i \frac{q_y}{2 M} \ \sum_f  e_f \ E^f _v (x,-q^2_\perp)  \ \right), 
\end{equation}
here $f= u, \ d $ are used for  up and down quarks,  $ H^f _v (x,-q^2_\perp)$,   $E^f _v (x,-q^2_\perp)$ denote the proton valence GPDs for unpolarized quark of flavor $f$, the transverse position vector is  $ {\bf b_\perp} = b_\perp \cos (\phi_{b_\perp}) \ \hat{e}_x + b_\perp \sin(\phi_{b_\perp})  \ \hat{e}_y $ and the transverse momentum transfer   is  ${\bf q_\perp} = q_\perp \cos (\phi_{q_\perp}) \ \hat{e}_x + q_\perp \sin(\phi_{q_\perp}) \ \hat{e}_y$  with a nucleon target of spin vector $\mathbf{S} = \cos (\phi_s) \ \hat{e}_x + \sin(\phi_s) \ \hat{e}_y$ such that $q_y = q_\perp \sin(\phi_{b_\perp}- \ \phi_{s})$.  The neutron transverse charge density is obtained utilizing isospin symmetry. Decomposing the GPDs in terms of their flavor components and evaluating the angular integrals for a nucleon polarized in the $+ x$ direction (i.e. $\phi_s = 0$), the proton transverse charge density becomes
\begin{multline}
\rho_T (x,\mathbf{b_\perp})_{proton} =  \int_0^{\infty} \frac{dq_\perp}{2 \pi} \ q_\perp  \ J_0 (b_\perp q_\perp ) \ \left[ e_u \ H^u_{v}(x,-q^2_\perp)  \ +  \ e_d \ H^d_{v}(x,-q^2_\perp) \right]  \\  -  \sin(\phi_{b_\perp}) \ \int_{0}^{\infty} \frac{dq_\perp}{2 \pi} \frac{q^2_\perp}{2 M} \  J_1 (b_\perp q_\perp) \left[ e_u \ E^u_v  (x,-q^2_\perp) \ + \ e_d \ E^d_v (x,-q^2_\perp) \right].
\end{multline}
From the above equation, the up and down quark densities   read
\begin{equation}
\begin{aligned}
\rho_u (x,\mathbf{b_\perp}) & =  \int_0^{\infty} \frac{dq_\perp}{2 \pi} \ q_\perp  \  J_0 (b_\perp q_\perp ) \ H^u_{v}(x,-q^2_\perp)  \ - \ \sin(\phi_{b_\perp})  \int_{0}^{\infty} \frac{dq_\perp}{2 \pi} \frac{q^2_\perp}{2 M} \  J_1 (b_\perp q_\perp)\ E^u_v  (x,-q^2_\perp) 
\label{eq:up-dens}
\end{aligned}
\end{equation}
\begin{equation}
\begin{aligned}
\rho_d (x,\mathbf{b_\perp}) & =  \int_0^{\infty} \frac{dq_\perp}{2 \pi} \ q_\perp  \  J_0 (b_\perp q_\perp ) \ H^d_{v}(x,-q^2_\perp)  \ - \ \sin(\phi_{b_\perp})  \int_{0}^{\infty} \frac{dq_\perp}{2 \pi} \frac{q^2_\perp}{2 M} \  J_1 (b_\perp q_\perp)\ E^d_v  (x,-q^2_\perp) 
\label{eq:down-dens}
\end{aligned}
\end{equation}
On the other hand, the electromagnetic potential corresponding to a transversly   polarized nucleon is given by~\cite{Tareq1}
\begin{multline}
A^{0\uparrow} \left(x, \mathbf{b_\perp} \right)  =  \frac{-1}{2 \pi}  \int_{0}^{\infty} \frac{dq_{\perp}}{q_\perp}  \left[- J_0 \left(b_{\perp} q_\perp \right) +  J_0 \left(b_0 q_\perp \right)  \right]\sum_f e_f \ H^f _v \left(x, - q^2_\perp \right)    +  \\  \frac{\sin\left( \phi_{b_\perp} -   \phi_s \right)}{4 \pi M}  \int_{0}^{\infty} dq_{\perp}  J_1 \left(b_{\perp} q_\perp \right)    \sum_f e_f \ E^f _v \left(x, - q^2_\perp \right) ,
\end{multline}
where $b_0$ is a reference point for the potential; in our case it is equal to the size of  the transverse charge density of the nucleon.
The field strengths corresponding to this potential read
\begin{multline}
F^{+i}(x,\mathbf{b_\perp}) = - \boldsymbol{\nabla} V \left( \mathbf{b_\perp} \right) = \left[ \frac{1}{2 \pi} \int_0^{\infty}  dq_{\perp}  J_1\left(b_\perp q_\perp \right)  \sum_f e_f \ H^f _v \left(x, - q^2_\perp \right)  -  \right. \\  \left. \frac{\sin(\phi_{b_\perp} - \phi_s)}{8 \pi M} \int_0^{\infty} q_{\perp} dq_{\perp} \left(  J_0 \left(b_\perp q_\perp \right) -  J_2 \left(b_\perp q_\perp \right) \right) \sum_f e_f \ E^f _v \left(x, - q^2_\perp \right)  \right]  \hat{b}_{\perp} \ -  \\   \frac{\cos(\phi_{b_\perp} - \ \phi_s)}{4 \pi M b_\perp} \int_0^{\infty}  dq_{\perp} J_1 \left(b_\perp q_\perp \right)\sum_f e_f \ E^f _v \left(x, - q^2_\perp \right)  \hat{\phi}_{b_\perp}, \label{eq:field}
\end{multline}
 the $x$ and $y$ components of $F^{+i}$ for proton  become after writing the polar unit vectors in terms of  the Cartesian ones 
\begin{multline}
F^{+x}(x, \mathbf{b_\perp})_{proton} =   \frac{\cos(\phi_{b_\perp})}{2 \pi} \int_0^{\infty}  dq_{\perp}  J_1\left(b_\perp q_\perp \right) \  \sum_f e_f \ H^f_v \left(x, - q^2_\perp \right) \ - \\  \frac{\cos(\phi_{b_\perp}) \sin(\phi_{b_\perp})}{8 \pi M} \int_0^{\infty} dq_{\perp} q_{\perp}  \left[  J_0 \left(b_\perp q_\perp \right) -  J_2 \left(b_\perp q_\perp \right) \right] \  \sum_f e_f  \ E^f _v \left(x, - q^2_\perp \right) \ + \\  \frac{\sin(\phi_{b_\perp}) \cos(\phi_{b_\perp})}{4 \pi M b_\perp} \int_0^{\infty}  dq_{\perp} J_1 \left(b_\perp q_\perp \right) \ \sum_f e_f \ E^f _v \left(x,-q^2_\perp \right) 
\end{multline}
\begin{multline}
F^{+y}(x, \mathbf{b_\perp})_{proton} =   \frac{\sin(\phi_{b_\perp})}{2 \pi} \int_0^{\infty}  dq_{\perp}  J_1\left(b_\perp q_\perp \right) \ \sum_f e_f \ H^f _v \left(x, - q^2_\perp \right)  \ -  \\   \frac{\sin^2(\phi_{b_\perp})}{8 \pi M} \int_0^{\infty}  dq_{\perp}  q_{\perp} \left[  J_0 \left(b_\perp q_\perp \right) -  J_2 \left(b_\perp q_\perp \right) \right]  \ \sum_f e_f \  E^f _v \left(x, - q^2_\perp \right) \ -  \\  \frac{\cos^2(\phi_{b_\perp})}{4 \pi M b_\perp} \int_0^{\infty}  dq_{\perp} J_1 \left(b_\perp q_\perp \right) \  \sum_f e_f \ E^f _v \left(x, - q^2_\perp \right),
\label{eq:Force-y-direction}
\end{multline}
the above forms allow to perform a flavor decomposition for the fields, for example the $y$-component of the field due to up  quarks  is
\begin{multline}
F^{+y}_{u}(x, \mathbf{b_\perp})_{proton} =   \frac{e_u \ \sin(\phi_{b_\perp})}{2 \pi} \int_0^{\infty}  dq_{\perp}  J_1\left(b_\perp q_\perp \right) \ H^u_v \left(x,- q^2_\perp \right)  \ -  \\   \frac{e_u \ \sin^2(\phi_{b_\perp})}{8 \pi M} \int_0^{\infty}  dq_{\perp}  q_{\perp} \left[  J_0 \left(b_\perp q_\perp \right) -  J_2 \left(b_\perp q_\perp \right) \right]  \   E^u _v \left(x, - q^2_\perp \right) \ -  \\  \frac{e_u \ \cos^2(\phi_{b_\perp})}{4 \pi M b_\perp} \int_0^{\infty}  dq_{\perp} J_1 \left(b_\perp q_\perp \right) \ E^u _v \left( x,- q^2_\perp \right),
\label{eq:Force-y-directionu}
\end{multline}
and due to down quarks is
\begin{multline}
F^{+y}_{d}( x,\mathbf{b_\perp})_{proton} =   \frac{e_d \ \sin(\phi_{b_\perp})}{2 \pi} \int_0^{\infty}  dq_{\perp}  J_1\left(b_\perp q_\perp \right) \ H^d _v \left( x,- q^2_\perp \right)  \ -  \\   \frac{e_d \ \sin^2(\phi_{b_\perp})}{8 \pi M} \int_0^{\infty}  dq_{\perp}  q_{\perp} \left[  J_0 \left(b_\perp q_\perp \right) -  J_2 \left(b_\perp q_\perp \right) \right]  \   E^d _v \left(x, - q^2_\perp \right) \ -  \\  \frac{e_d \ \cos^2(\phi_{b_\perp})}{4 \pi M b_\perp} \int_0^{\infty}  dq_{\perp} J_1 \left(b_\perp q_\perp \right) \ E^d _v \left(x, - q^2_\perp \right).
\label{eq:Force-y-directiond}
\end{multline}
As we shall see, the above decomposition of the nucleon's fields motivates to define the $q \gamma q$ correlator for each flavor in terms of the  field  due to the corresponding flavor.

\section{$q \gamma q$ Correlator in Inclusive DIS}

 \noindent The  $q \gamma q$  correlator (denoted by $F^q_{FT}(x,x)$)  as defined in Eq.(10) in Ref.\cite{Metz1} is given by
 \begin{equation}
\int \frac{d \xi^{-} d \zeta^{-}}{2 (2 \pi)^2} e^{ixP^{+} \xi^{-}} \langle P,S | \bar{\psi}^q (0) \gamma^+ \ e \  F^{+i}_{QED}(\zeta)  \psi^q(\xi) |P,S \rangle = - M \epsilon^{ij}_T S^j_T F^q_{FT}(x,x)
\label{eq:FFTqed}
\end{equation}
here $ F^{+i}_{QED}(\zeta)$ is the electromagnetic field strength tensor. The above definition follows from an  analogy with the so called $QCD$ soft gluon pole matrix element (Qiu-Sterman function) $T^q_F(x,x')$ which is defined through\cite{Qiu91, Qiu98}
\begin{equation}
\int \frac{d \xi^{-} d \zeta^{-}}{4 \pi} e^{ixP^{+} \xi^{-}} \langle P,S | \bar{\psi}^q (0) \gamma^+   F^{+i}_{QCD}(\zeta)  \psi^q(\xi) |P,S \rangle = -  \epsilon^{ij}_T  S^j_T  T^q_{F} (x,x)
\label{eq:FFTqcd}
\end{equation}
where  $F^{+i}_{QCD}$ is the $QCD$ field strength tensor and the corresponding  matrix element represents the average force of the ejected quark in a $SIDIS$ process~\cite{Burkardt-transv-force}, i.e. in the final state interaction. In the same way the matrix element in  Eq\eqref{eq:FFTqed} is interpreted  as the average transverse force on the electron in IDIS process due to initial and final state interactions~\cite{BurkardtFFT}. Based on the above analogy, and using the results of the previous section for the flavor decomposition of the fields of a transversely polarized nucleon, we  define the $q \gamma q$ correlator for each flavor in terms of the corresponding field component for that flavor such that
\newline
 \begin{equation}
  \int \frac{d \xi^{-} d \zeta^{-}}{4 \pi} e^{ixP^{+} \xi^{-}} \langle P,S | \bar{\psi}^q (0) \gamma^+ \ e \  F^{+i}_{q/N} (x,\zeta)  \psi^q(\xi) |P,S \rangle = - M \epsilon^{ij}_T  S^j_T  \mathcal{N}_{q/N}(x)   F^{q/N}_{FT}(x,x)
\label{eq:FFTTP1}
\end{equation}
where 
\vspace{4mm}
$$\mathcal{N}_{q/N}(x) = (2 \pi)^2 e^2_q \  \rho_{q/N}(x)$$
and  
$e=\sqrt{4 \pi  \alpha_{em}}$, $\alpha_{em}=1/137$, $M=0.938 \ GeV$, $\epsilon^{12}_T=1$, $S^x_T=1$ and  the quark densities $\rho_{q/N}(x,\mathbf{b}_\perp)$ are give in Eqs.(\ref{eq:up-dens},\ \ref{eq:down-dens}) .
The function $\mathcal{N}_{q/N}(x)$ is consistent with the results obtained for the $q g q$ correlator using  Sivers function extracted from $SIDIS$ data~\cite{Anselmino2005} and also with the calculation of the  T-odd $q g q$ correlation functions in the diquark model~\cite{Shmidt2012} and with the parametrization of the Qiu-Sterman function $T^q_{F}(x,x)$ from $pp^{\uparrow}$ data\cite{Couvaris}, where in all of the above, similar normalization function is usually used. Notice the main difference between Eqs.(\ref{eq:FFTqed},\ref{eq:FFTqcd}) and Eq.\eqref{eq:FFTTP1}; in Eq.\eqref{eq:FFTqcd}, the matrix element of  $F^{+i}_{QCD}$ represents the average force experienced by the active quark from the nucleon remnants, i.e. in the FSI~\cite{Burkardt-transv-force}. Similarly, in Eq.\eqref{eq:FFTqed}, the matrix element of $F^{+i}_{QED}$ is the average force experienced by the electron when passing through the nucleon in the FSI and ISI~\cite{BurkardtFFT}. On the other hand, in Eq.\eqref{eq:FFTTP1},  $F^{+i}_{q/N}$ is the field experienced by the electron in IDIS process from quarks of flavor $q$. Therefore, the $q \gamma q$ correlator corresponding to flavor $q$ can be expressed as
\begin{equation}
F^{q/N}_{FT}(x,x) \ = \ \frac{-1}{ M  S^{i}_T \epsilon^{ij}_T  \mathcal{N}_{q/N}(x)} \int d^2 \mathbf{b}_\perp \ \rho_{q/N}(x,\mathbf{b}_\perp) \ e \ F^{+i}_{q/N} (x,\mathbf{b}_\perp),
\label{eq:FFTTP2}
\end{equation}
note that the only fields contribute to the above formula are $ F^{+y}_{q/N} (x,\mathbf{b}_\perp) $; since the verage transverse momentum in the $x$-direction vanishes by symmetr~\cite{Tareq1}. The figures below are plots of $x F^{q/N}(x,x)$  ($N=proton \ or \  neutron$),  using three approachs, the impact parameter transverse potential, and two models for the correlator $F^{q/N}_{FT}(x,x)$, Burkardt's model from Ref.\cite{BurkardtFFT} and Metz's et al. model from Ref.\cite{Metz1}. In the  calculations of  $F^{q/N}_{FT}(x,x)$ using Burkardt and Metz models, the parametrization of the Qui-Sterman functions $T^q_F(x,x)$ used in these models was taken from Ref.\cite{Qui-Sterman-TF2011}, this parametrization is based on Sivers functions parametrization which is taken from Ref.\cite{Anselmino2009}. In the calculation of  $F^{q/N}_{FT}(x,x)$  using the transverse nucleon potential, the generalized parton distribution (GPD) parametrization was taken from Ref.~\cite{GPD2} and the transverse radii of the   charge densities for proton and neutron were taken from Ref.~\cite{CarlsonVanderhaeghen1}.
\begin{figure}[H]
\begin{center}
\includegraphics[scale=0.93]{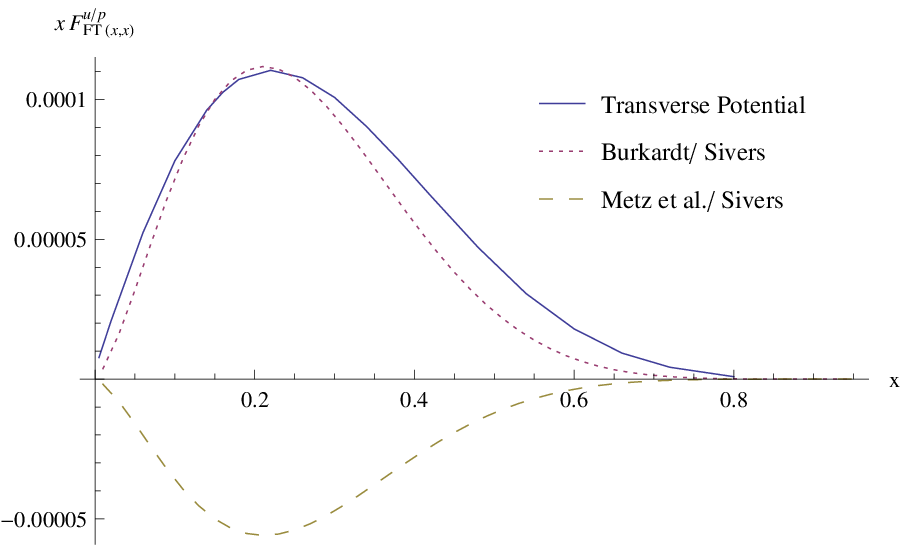}
\includegraphics[scale=0.85]{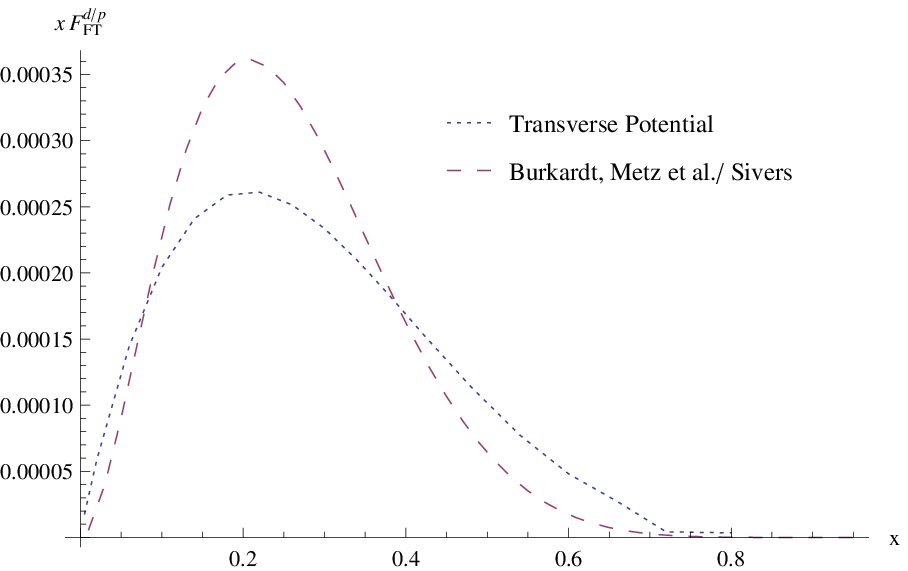}
\caption{ $x$ times the correlator  $F^{q/p}_{FT}(x,x)$ as a function of  $x$ for proton using the impact parameter transverse potential from Ref.~\cite{Tareq1}, Burkardt's model from Ref.\cite{BurkardtFFT} and Metz's model from Ref.\cite{Metz1}.}
 \label{fig:FFT-p}
\end{center}
\end{figure}

\begin{figure}[H]
\begin{center}
\includegraphics[scale=0.93]{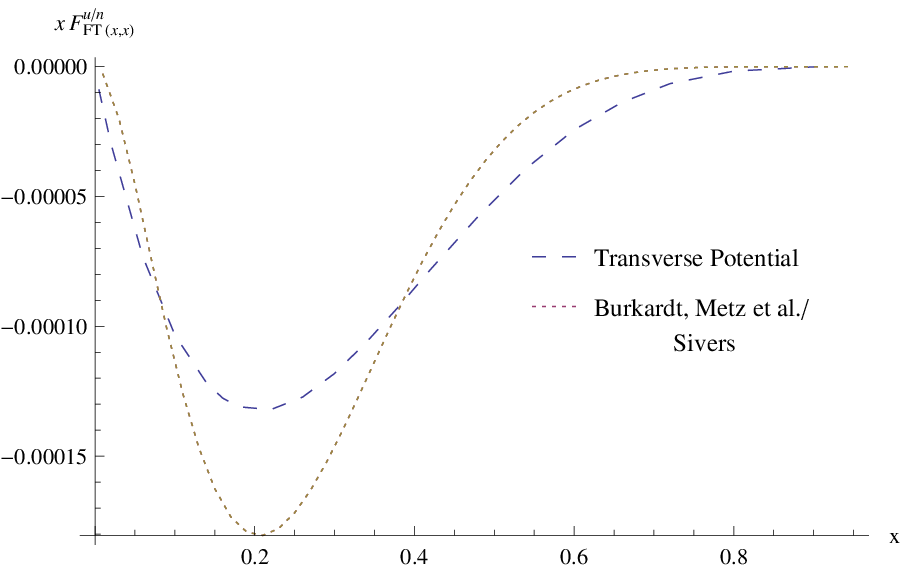}
\includegraphics[scale=0.85]{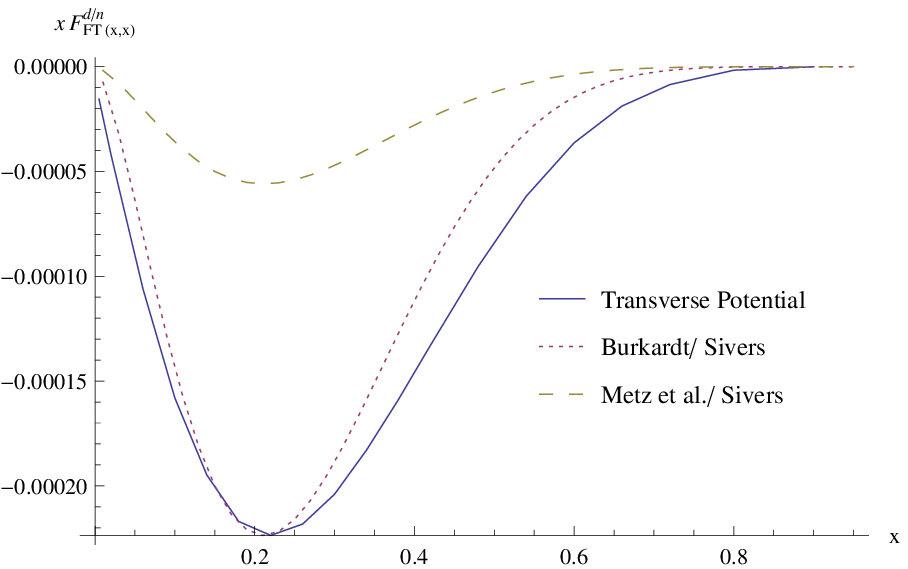}
\caption{ $x$ times the correlator  $F^{q/n}_{FT}(x,x)$ as a function of  $x$ for neutron using the impact parameter transverse potential from Ref.~\cite{Tareq1}, Burkardt's model from Ref.\cite{BurkardtFFT} and Metz's model from Ref.\cite{Metz1}.}
 \label{fig:FFT-n}
\end{center}
\end{figure}
\section{Calculation of the Transverse Target Single Spin Asymmetry in IDIS}
\begin{figure}[H]
\begin{center}
\includegraphics{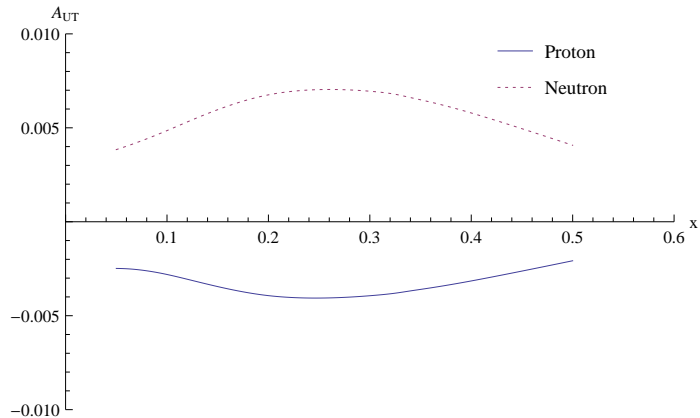}
\caption{Proton and neutron transverse target asymmetry at $y=0.6$ \ and \ $Q^2=1 \ GeV^2$, corresponding to typical Jefferson Lab kinematics.}
 \label{fig:FFT-tilda}
\end{center}
\end{figure}

In this section we will use the results obtained in the previous section for the correlator $F^{q/N}_{FT}(x,x)$ to calculate the SSA in inclusive DIS for transversely polarized proton and neutron. In Ref.~\cite{Metz1} the transverse target SSA is given by
\begin{equation}
A^N_{UT} = - \frac{2 \pi M}{Q} \frac{2-y}{\sqrt{1-y}} \frac{\sum_q e^2_q \ x \ \tilde{F}^{q/N}_{FT}(x,x)}{\sum_q e^2_q  \ x \ f^{q/N}_{1}(x)},
\label{eq:TTSSA-Metz}
\end{equation}
where 
\begin{equation}
\tilde{F}_{FT} = F_{FT}(x,x)  - x \frac{d}{dx}F_{FT}(x,x),
\label{FFT-tilda}
\end{equation}
and $f^q_1$ is the unpolarized quark distribution function for a quark of flavor $q$ which are taken from Ref.~\cite{GPD2}. The relation between the variables in \eqref{eq:TTSSA-Metz} is given by $y=Q^2/(xs)$, with $s$ denotes the square of the center of mass energy and $y$ is the fraction of the electron energy carried by the virtual photon. Figure \ref{fig:FFT-tilda} is a plot for proton and neutron transverse target asymmetry for typical Jefferson Lab kinematics at $y=0.6$ and $Q^2=1 \ GeV^2$.

\section{Conclusion}
We calculated the $q \gamma q$   correlator $F^{q/n}_{FT}(x,x)$ associated with a transversly polarized nucleon in IDIS process by direct evaluation of the corresponding matrix element using the electromagnetic potential for a transversely polarized nucleon. The results are compared with the two existing models for the $q \gamma q$   correlator. Clearly from the above figures, Burkardt's and Metz's models agree in the $q \gamma q$ correlator for  proton and neutron minority flavors but disagree in sign and magnitude for  proton up quarks while  for neutron down quarks, the two models significantly disagree  in magnitude but agree in sign. On the other hand, the results obtained using the transverse nucleon potential are in agreement in sign and magnitude with Burkardt's model for  proton and  neutron majority flavors. For the minority flavors,  there is agreement in sign, but one notice a slight difference in  magnitude for for both proton and neutron. However, the results obtained  for  $F^{q/n}_{FT}(x,x)$  using the transverse potential  are in better agreement with the sum rule conecting the correlators of the two flavors~\cite{Burkardt2014}.
\vspace{3mm}

\noindent Using the calculated $q \gamma q$ correlator, we estimated the transverse target SSA for inclusive DIS, the neutron results are consistent with recent calculations based on Sivers function parametrization~\cite{Metz1}. For the proton, the SSA results are almost half those for neutron, which is a main difference from the above reference and other calculations~\cite{Afanasev-DIS}, however the sign of our calculations is consistent with the above references.

\vspace{5mm}
\bibliographystyle{unsrt}
\bibliography{A_UT.bib}

\begin{thebibliography}{10}

\bibitem{Qiu91}
Jianwei Qiu and George Sterman.
\newblock Single transverse spin asymmetries in direct photon production.
\newblock {\em Nuclear Physics B}, 378(1–2):52 -- 78, 1992.

\bibitem{Qiu98}
Jianwei Qiu and George Sterman.
\newblock Single transverse-spin asymmetries in hadronic pion production.
\newblock {\em Phys. Rev. D}, 59:014004, Nov 1998.

\bibitem{Shmidt2012}
Zhun Lu and Ivan Schmidt.
\newblock T-odd quark–gluon–quark correlation function in the diquark
  model.
\newblock {\em Physics Letters B}, 712(4–5):451 -- 455, 2012.

\bibitem{Q-g-correlator}
Zhong-Bo Kang, Jian-Wei Qiu, and Hong Zhang.
\newblock Quark-gluon correlation functions relevant to single transverse spin
  asymmetries.
\newblock {\em Phys. Rev. D}, 81:114030, Jun 2010.

\bibitem{Couvaris}
Chris Kouvaris, Jian-Wei Qiu, Werner Vogelsang, and Feng Yuan.
\newblock Single transverse-spin asymmetry in high transverse momentum pion
  production in $pp$ collisions.
\newblock {\em Phys. Rev. D}, 74:114013, Dec 2006.

\bibitem{Boer2015}
Dani{\"e}l Boer.
\newblock Average transverse momentum quantities approaching the lightfront.
\newblock {\em Few-Body Systems}, 56(6):439--445, 2015.

\bibitem{Metz1}
A.~Metz, D.~Pitonyak, A.~Sch\"afer, M.~Schlegel, W.~Vogelsang, and J.~Zhou.
\newblock {\em Phys. Rev. D}, 86:094039, Nov 2012.

\bibitem{BurkardtFFT}
Matthias Burkardt.
\newblock Quark orbital angular momentum and final state interactions.
\newblock In {\em International Journal of Modern Physics: Conference Series},
  volume~37, page 1560035. World Scientific, 2015.

\bibitem{Tareq1}
Tareq Alhalholy and Matthias Burkardt.
\newblock Impact parameter dependent potentials and average transverse momentum
  in inclusive dis.
\newblock {\em Phys. Rev. D}, 93:125019, Jun 2016.

\bibitem{BurkardtGPDs}
Matthias Burkardt.
\newblock {\em International Journal of Modern Physics A}, 18(02):173--207,
  2003.

\bibitem{Burkardt-transv-force}
Matthias Burkardt.
\newblock Transverse force on quarks in deep-inelastic scattering.
\newblock {\em Physical Review D}, 88(11):114502, 2013.

\bibitem{Anselmino2005}
M.~Anselmino, M.~Boglione, U.~D'Alesio, A.~Kotzinian, F.~Murgia, and
  A.~Prokudin.
\newblock Extracting the sivers function from polarized semi-inclusive deep
  inelastic scattering data and making predictions.
\newblock {\em Phys. Rev. D}, 72:094007, Nov 2005.

\bibitem{Qui-Sterman-TF2011}
Zhong-Bo Kang, Jian-Wei Qiu, Werner Vogelsang, and Feng Yuan.
\newblock Observation concerning the process dependence of the sivers
  functions.
\newblock {\em Phys. Rev. D}, 83:094001, May 2011.

\bibitem{Anselmino2009}
M.~Anselmino, M.~Boglione, U.~D'Alesio, A.~Kotzinian, S.~Melis, F.~Murgia,
  A.~Prokudin, and C.~T{\"u}rk.
\newblock Sivers effect for pion and kaon production in semi-inclusive deep
  inelastic scattering.
\newblock {\em The European Physical Journal A}, 39(1):89--100, 2009.

\bibitem{GPD2}
M.~Guidal, M.~V. Polyakov, A.~V. Radyushkin, and M.~Vanderhaeghen.
\newblock {\em Phys. Rev. D}, 72:054013, Sep 2005.

\bibitem{CarlsonVanderhaeghen1}
Carl~E. Carlson and Marc Vanderhaeghen.
\newblock {\em Phys. Rev. Lett.}, 100:032004, Jan 2008.

\bibitem{Burkardt2014}
Matthias Burkardt.
\newblock Quark orbital angular momentum and final state interactions.
\newblock In {\em International Journal of Modern Physics: Conference Series},
  volume~25, page 1460029. World Scientific, 2014.

\bibitem{Afanasev-DIS}
A.~Afanasev, M.~Strikman, and C.~Weiss.
\newblock Transverse target spin asymmetry in inclusive dis with two-photon
  exchange.
\newblock {\em Phys. Rev. D}, 77:014028, Jan 2008.

\end{thebibliography}

\end{document}